\documentclass[12pt,thmsa]{article}

\usepackage{amsfonts}
\usepackage{graphicx}
\usepackage{epsfig}
\usepackage{graphics}

\makeatletter
\newcommand{\row}[1]%
{\mathord{\buildrel{\lower3pt%
\hbox{$\scriptscriptstyle\rightarrow$}}\over #1}}

\newcommand{\dyadic}[1]{\mathord{\dyadic@rrow{#1}}}
\newcommand{\dyadic@rrow}[1]{
\begin{picture}(12,12)(-1,0)
\put(-1,9){\makebox(0,0)[t]{$\scriptscriptstyle\downarrow$}}
\put(-1,9){\makebox(0,0)[l]{$\scriptscriptstyle\longrightarrow$}}
\put(5,0){\makebox(0,0)[b]{$#1$}}
\end{picture}
}

\newcommand{\bra}[1]{\bigl\langle #1 \bigr|}
\newcommand{\ket}[1]{\bigl| #1 \bigr\rangle}

\begin{document}

\begin{center}
{\Large Enhancing   entanglement, local and non-local  information of accelerated  two-qubit and two- qutrit systems via  weak-reverse measurements }\\
 {N. Metwally\\}
$^1$ Department of Mathematics, College of Science, Bahrain
University, Bahrain \\
$^2$Department of Mathematics, Faculty of Science,
Aswan University, Aswan, Egypt \\
email:nmetwally@gmail.com, nmetwally@uob.edu.bh \\
Tel:00973 33237474, ~ Fax: 00973 17449145
\end{center}
\date{\today }

\begin{abstract}
The possibility of recovering and protecting  the entanglement of
accelerated 2-qubit and 2-qutrit  systems is discussed  using
weak-reverse measurements. The accelerated  partial entangled
states are more  responsive to be protected than the accelerated
maximum entangled states. The accelerated coded  local information
in qutrit system  is more robust than that encoded in a 2-qubit
system, and it  can be conserved  even for larger values of
accelerations. Meanwhile, the non-accelerated information in qubit
system  is not affected by the local operation compared with that
depicted on qutrit system. For larger values of accelerations, the
weak-reverse measurements can improve the coherent information at
the expanse of the accelerated information.

\end{abstract}

\section{Introduction}

 There are some techniques that have been used to recover  the losses of entanglement
and  protect it from decoherence. Among of these techniques is the
quantum purification
\cite{Bennt,Deutsch,Metwally2002,Metwally2006}. Moreno et.
al.,\cite{Moreno} showed that, it is possible to improve the
coherence by using nested environments. Xiao et. al
\cite{Xiao2016} displayed  that the teleportation of Fisher
information can be enhanced by using partial measurements. The
possibility of purifying and concentrating entanglement by using
local filtering is discussed by Yashodamma et. al, \cite{Yas2014}.
The immunity and protecting the entanglement of accelerated
qubit-qutrit system by using the local filtering technique is
investigated by Metwally\cite{Metwally2016-0}. Recently, it has
been shown that the local weak-reverse measurements,(WRM)  can be
used as a technique to protect and improve the correlation in
qubit-qutrit systems \cite{Liang}. Xiao et. al.\cite{Xiao}, showed
that, the entanglement losses  caused by amplitude damping
coherence can be retrieved by using the WRM.

It is well known that, the accelerated systems lose some of its
correlations, and consequently, their efficiencies to perform
quantum information tasks
decrease\cite{Metwally2013,Downes,Sagheer}. This decay depends on
the value of the acceleration and the dimensional of the
accelerated system\cite{Metwally2016}. Therefore we are motivated
to investigate the possibility of using the weak-reverse
measurement's technique to improve some properties of accelerated
system. The main task of this contribution is   quantifying the
amount of entanglement between the two partners in the presence of
weak-reverse measurements, investigating the dynamics of the
accelerated/non accelerated information ( which is encoded on the
accelerated/non accelerated particle) and the coherent information
between the partners.

 The outline of this manuscript is described as follows: The suggested protocol is described in Sec.2.
  The suggested system and its
analytical solutions are given in Sec.$3$. The effect of
 the weak-reverse'strengths on the   degree of entanglement is
 discussed in Sec.$4$., where  different initial states settings are
 considered. The dynamics of the accelerated, non-accelerated and
 coherent information is investigated in  Sec.$5$. Finally, a
conclusion is given  in Sec.$6$.

\section{Model}

Our idea is  to consider two partners, Alice and Bob, who share a
two-qubit or two-qutrit systems.  These systems are initially
prepared in maximum or partial entangled states. Here, we consider
only Alice's particle is accelerated, which may be a qubit or a
qutrit. The partners decided in advance to use the weak-reverse
measurement to protect their quantum communication channel. In the
following steps, we describe the proposed protocol:

\begin{enumerate}
\item{Weak measurements:\\} Here, both users Alice and Bob perform
the weak measurements on their own particles, either qubit or
qutrit.  We assume that, the two partners share initially a state
defined  by $\rho_{ab}^{ini}$.  After performing the weak
measurements, the partners share a new state defined by
$\rho_{ab}^{out}$,
\begin{equation}
\rho_{ab}^{out}=\mathcal{W}_j\mathcal{W}_j\left(\rho_{ab}^{ini}\right)\mathcal{W}_j^{\dagger}\mathcal{W}_j^{\dagger},
~i,j=q,t,
\end{equation}
where $\mathcal{W}_i$ are the weak measurements which are defined
by the following operators \cite{kim,Xiao2013}
\begin{eqnarray}\label{Wm}
\mathcal{W}_q&=&\ket{0}\bra{0}+\sqrt{1-\alpha_q^{(\ell)}}~\ket{1}\bra{1}~\quad\mbox{(for qubit)},\nonumber\\
\mathcal{W}_t&=&\ket{0}\bra{0}+\sqrt{1-\alpha_t^{(1)}}~\ket{1}\bra{1}+\sqrt{1-\alpha_t^{(2)}}~\ket{2}\bra{2}~\quad\mbox{(for-qutrit)},
\end{eqnarray}
and $\alpha_q^{(\ell)}$, $\alpha_t^{(\ell)}, \ell=1,2$ are the
strengths of the weak measurements of the qubit and the qutrit,
respectively.

 \item{Acceleration step:\\}
It is assumed that, only  Alice' particle is moving with a uniform
acceleration  meanwhile Bob's partial is assumed to be inertial.
If the shared particles are of fermions types, then in the
Minkowski frame  the qubit system is transformed in the Rindler
frame as,
\begin{eqnarray}
\ket{0_M}&=&\cos r\ket{0}_I\ket{0}_{II}+\sin
r\ket{1}_I\ket{1}_{II},\quad \ket{1_M}=\ket{1}_{I}\ket{0}_{II}
\end{eqnarray}
while, for the qutrit system, the vacuum, the spin up and spin
down states  are transformed into  Rindler space as,
\begin{eqnarray}
\ket{0_M}&=&\cos^2r\ket{0}_{I}\ket{0}_{II}+e^{i\phi}\sin r\cos
r(\ket{\mathcal{U}}_{I}\ket{\mathcal{D}}_{II}+\ket{\mathcal{D}}_{I}\ket{\mathcal{U}}_{II})
+e^{2i\phi}sin^2r\ket{\mathcal{D}}_{I}\ket{\mathcal{P}}_{II},
\nonumber\\
 \ket{\mathcal{U}_M}&=&cos
r\ket{\mathcal{U}}_I\ket{0}_{II}+e^{i\phi}\sin
r\ket{\mathcal{P}}_I\ket{\mathcal{U}}_{II},
\nonumber\\
\ket{\mathcal{D}_M}&=&cos
r\ket{\mathcal{D}}_I\ket{0}_{II}-e^{i\phi}\sin
r\ket{\mathcal{P}}_I\ket{\mathcal{D}}_{II},
\end{eqnarray}
where $\ket{\mathcal{U}}, \ket{\mathcal{D}}$ and
$\ket{\mathcal{P}}$ are the spin up, spin down and pair states,
respectively. The acceleration $r$ is defined such that $\tan
r=Exp[-\pi\omega\frac{c}{a}]$, ~$0\leq r\leq \pi/4$, $-\infty\leq
a\leq\infty$, $\omega$ is the frequency, $c$ is the speed of light
and $\phi$ is the phase space which can be absorbed in the
definition of the operators \cite{Jason2013}. After tracing the
particle in the second region (II), the final state represents the
accelerated quantum channel between Alice and Bob,
$\rho_{ab}^{acc}$.

 \item {Reverse measurement step:\\}
In this step, the users apply the reverse measurement operations
on the accelerated state  $\rho_{ab}^{acc}$ to obtain the final
state as,

\begin{equation}\label{Final}
\rho_{ab}^{Final}=\mathcal{R}_i\mathcal{R}_i\left(\rho_{ab}^{acc}\right)\mathcal{R}^{\dagger}_{i}\mathcal{R}^\dagger_{i},\quad
i=q,t,
\end{equation}
where
\begin{eqnarray}\label{Rm}
\mathcal{R}_q&=&\sqrt{1-\beta_q^{(\ell)}}~\ket{0}\bra{0}+\ket{1}\bra{1},\quad
\ell=1,2
\nonumber\\
\mathcal{R}_t&=&\sqrt{(1-\beta_t^{(1)})(1-\beta_t^{(2)})}~\ket{0}\bra{0}+\sqrt{1-\beta_t^{(1)}}~\ket{1}\bra{1}
+\sqrt{1-\beta_t^{(2)}}~\ket{2}\bra{2},
\end{eqnarray}
\end{enumerate}
and $\beta_q^{(\ell)}$, $\beta_t^{(\ell)}, \ell=1,2$ are the
strengths of the reverse measurement operations.
  Some  properties  of  the final state
Eq.(\ref{Final}), such as  the behavior of entanglement,
accelerated local  non-local information are examined.
Particularly,  the effect of the initial state  settings and the
local operations strengths  on these properties.

\subsection{The suggested Systems}
\begin{enumerate}
\item {\bf Two-qubit system\\} In this section, it is assumed that
the partners, Alice and Bob, share a partially entangled state,
which is called $X-$ state. This state can be written, using the
computational basis as,

\begin{eqnarray}\label{x-state}
\rho_x&=&\mathcal{B}_1\left(\ket{00}\bra{00}+\ket{11}\bra{11}\right)+\mathcal{B}_2\left(\ket{00}\bra{11}+\ket{11}\bra{00}\right)
\nonumber\\
&+&\mathcal{B}_3\left(\ket{01}\bra{01}+\ket{10}\bra{10}\right)+\mathcal{B}_4\left(\ket{01}\bra{10}+\ket{10}\bra{01}\right)
\end{eqnarray}
where $\mathcal{B}_1=\frac{1}{4}(1+c_{33}),
\mathcal{B}_2=\frac{1}{4}(c_{11}+c_{22}),
\mathcal{B}_3=\frac{1}{4}(1-c_{33}), $ and
$\mathcal{B}_4=\frac{1}{4}(c_{11}-c_{22})$, and $c_i,i=1,2,3$ are
the diagonal  elements of $3\times 3$ cross dyadic. The state
(\ref{x-state}) can be described by its  Bloch vectors,
$\row{s}_j=0,j=1,2$ and  a dyadic $\dyadic{C}$ with $c_{ij}=0$ for
$i\neq j$. From this state, one can get maximum entangled classes,
MES. For example, if we set $c_{11}=c_{22}=c_{33}=-1$, one gets
the singled state, $\rho_{\psi^-}=\ket{\psi^-}\bra{\psi^-}$. The
Werener state can be obtained if we set $c_{11}=c_{22}=c_{33}=-x$
, etc. As mentioned earlier, only Alice's particle will be
accelerated \cite{MetwallyJMPB}. Due to the acceleration the
degree of entanglement decreases. Therefore, the partners try to
improve the degree of entanglement by applying the weak-reverse
measurements. At the end of the protocol, the partners share the
state.

\begin{eqnarray}\label{Final-qubit}
\rho_{ab}^{Final}&=&\frac{1}{\mathcal{N}_q}\Bigl\{\mathcal{B}_1\ket{00}\bra{00}+\tilde\mathcal{B}_2\ket{00}\bra{11}
+\tilde\mathcal{B}_3\ket{01}\bra{01}+\tilde\mathcal{B}_4\ket{01}\bra{10}
\nonumber\\
&+&\mathcal{B}_5\ket{10}\bra{10}+\tilde\mathcal{B}_6\ket{10}\bra{01}
+\tilde\mathcal{B}_7\ket{11}\bra{11}+\tilde\mathcal{B}_8\ket{11}\bra{00}\Bigr\}
\end{eqnarray}

where,
\begin{eqnarray}
\tilde\mathcal{B}_1&=&c_1^2\mathcal{B}_1(1-\beta_q^{(1)})(1-\beta_q^{(2)}),\quad
\nonumber\\
\tilde\mathcal{B}_2&=&c_1\mathcal{B}_2\sqrt{1-\beta_q^{(1)}}\sqrt{1-\beta_q^{(2)}}\sqrt{1-\alpha_q^{(1)}}\sqrt{1-\alpha_q^{(2)}},
\nonumber\\
\tilde\mathcal{B}_3&=&c_1^2\mathcal{B}_3(1-\beta_q^{(1)})(1-\alpha_q^{(2)}),\quad
\nonumber\\
\tilde\mathcal{B}_4&=&c_1\mathcal{B}_4\sqrt{1-\beta_q^{(1)}}\sqrt{1-\beta_q^{(2)}}\sqrt{1-\alpha_q^{(1)}}\sqrt{1-\alpha_q^{(2)}},
\nonumber\\
\tilde\mathcal{B}_5&=&(1-\beta_q^{(2)})\left[s_1^2\mathcal{B}_1+(1-\alpha_q^{(1)})\mathcal{B}_3\right],\quad
\tilde\mathcal{B}_6=\tilde\mathcal{B}_4
\nonumber\\
\tilde\mathcal{B}_7&=&(1-\alpha_q^{(1)})(1-\alpha_q^{(2)})
 \mathcal{B}_1,~\quad
\tilde\mathcal{B}_8=\tilde\mathcal{B}_2,
\end{eqnarray}
and
$\mathcal{N}_q=\tilde\mathcal{B}_1+\tilde\mathcal{B}_3+\tilde\mathcal{B}_5+\tilde\mathcal{B}_6$
is the normalization factor and $c_1=\cos r, s_1=\sin r$.

\item{Two-qutrit system:\\} In this section, we investigate the
effect of the local-weak-reverse operations on a system consists
of two qutrits, where it is assumed that only Alice's qutrit will
be accelerated. In the computational basis, the initial sate of
this system can be written as \cite{Parisio},
\begin{equation}
\ket{\psi_{t}}=\frac{1}{\sqrt{2+\gamma^2}}\left(\ket{00}+\ket{11}+\gamma\ket{22}\right),
\end{equation}
where it turns into a maximum entangled state for $\gamma=1$. The
users apply the  protocol which is described in Sec.2. At the end
the partners share the following state,
\begin{eqnarray}\label{Final-qutrit}
\rho_{ab}^{Final}&=&\frac{1}{\mathcal{N}_t}\Bigl\{\mathcal{D}_1\ket{00}\bra{00}+\mathcal{D}_2\ket{00}\bra{11}+\mathcal{D}_3\ket{10}\bra{10}+
\mathcal{D}_4\ket{11}\bra{00}+\mathcal{D}_5\ket{11}\bra{11}
\nonumber\\
&&+\mathcal{D}_6\ket{20}\bra{20}+\mathcal{D}_7\ket{22}\bra{00}+\mathcal{D}_8\ket{22}\bra{11}+\mathcal{D}_9\ket{22}\bra{22}+\mathcal{D}_{10}\ket{00}\bra{22}
\nonumber\\&& +\mathcal{D}_{11}\ket{11}\bra{22}\Bigr\},
\end{eqnarray}
where
\begin{eqnarray}
 \mathcal{D}_1&=&c_1^2\mathcal{R}_{00}^2\mathcal{A}_1, \quad
 \mathcal{D}_2=c_1^3\mathcal{R}_{00}\mathcal{R}_{11}\mathcal{A}_2,
 \quad \mathcal{D}_3=c_1^2s_1^2\mathcal{R}_{10}^2\mathcal{A}_1,
 \nonumber\\
 \mathcal{D}_4&=&c_1^3\mathcal{R}_{00}\mathcal{R}_{11}\mathcal{A}_4,\quad
  \mathcal{D}_5=c_1^3\mathcal{R}_{11}^2\mathcal{A}_5,\quad
  \mathcal{D}_6=c_1^2s_1^2\mathcal{R}_{20}^2\mathcal{A}_1,
  \nonumber\\
 \mathcal{D}_7&=&c_1^3\mathcal{R}_{22}\mathcal{R}_{00}\mathcal{A}_7,\quad
  \mathcal{D}_8=c_1^2\mathcal{R}_{22}\mathcal{R}_{11}\mathcal{A}_8,\quad
  \mathcal{D}_9=c_1^2\mathcal{R}_{22}^2\mathcal{A}_9,
  \nonumber\\
  \mathcal{D}_{10}&=&\mathcal{R}_{00}\mathcal{R}_{22}\mathcal{A}_3c^3,\quad
 \mathcal{D}_{11}=\mathcal{R}_{11}\mathcal{R}_{22}\mathcal{A}_6c^2,\quad
\end{eqnarray}
The normalization factor is given by
$\mathcal{N}_t=\mathcal{D}_1+\mathcal{D}_3+\mathcal{D}_5+\mathcal{D}_6+\mathcal{D}_9$
and the coefficients $\mathcal{A}_i, i=1..9$ are given by,
 \begin{eqnarray*}
\mathcal{A}_1&=&\frac{1}{2+\gamma^2},~\quad\quad
\mathcal{A}_2=\frac{\sqrt{1-\alpha_t^{(1)}}\sqrt{1-\alpha_t^{(2)}}}{2+\gamma^2},\quad
\mathcal{A}_3=\frac{\gamma\sqrt{1-\alpha_t^{(1)}}\sqrt{1-\alpha_t^{(2)}}}{2+\gamma^2},~\quad\quad
\nonumber\\
\mathcal{A}_4&=&\mathcal{A}_2,~\quad\quad
\mathcal{A}_5=(2+\gamma^2)\mathcal{A}_2^2,~\quad\quad
\mathcal{A}_6=\gamma\mathcal{A}_2\sqrt{1-\alpha_t^{(1)}}\sqrt{1-\alpha_t^{(2)}},\quad
\nonumber\\
\mathcal{A}_7&=&\frac{\gamma}{2+\gamma^2}\sqrt{1-\alpha_t^{(1)}}\sqrt{1-\alpha_t^{(2)}},~\quad
\mathcal{A}_8=\frac{\gamma(1-\alpha_t^{(1)})}{2+\gamma^2}\sqrt{1-\alpha_t^{(2)}}\sqrt{1-\alpha_t^{(2)}},\quad
\nonumber\\
\mathcal{A}_9&=&\frac{\gamma^2}{2+\gamma^2}(1-\alpha_t^{(1)})(1-\alpha_t^{(2)}),
 \end{eqnarray*}
and  $\mathcal{R}_{ij},i,j=0,1,2$ are given by
\begin{eqnarray*}
\mathcal{R}_{00}&=&\sqrt{(1-\beta_1^{(1)})(1-\beta_2^{(1)})}\sqrt{(1-\beta_1^{(2)})(1-\beta_2^{(2)})},\quad
\nonumber\\
\mathcal{R}_{01}&=&\sqrt{1-\beta_1^{(2)}}\sqrt{(1-\beta_1^{(1)})(1-\beta_2^{(1)})},\quad
\nonumber\\
\mathcal{R}_{02}&=&\sqrt{1-\beta_2^{(2)}}\sqrt{(1-\beta_1^{(1)})(1-\beta_2^{(1)})},\quad
\nonumber\\
\mathcal{R}_{10}&=&\sqrt{1-\beta_1^{(1)}}\sqrt{(1-\beta_1^{(2)})(1-\beta_2^{(2)})},
\nonumber\\
\mathcal{R}_{11}&=&\sqrt{1-\beta_1^{(1)}}\sqrt{1-\beta_1^{(2)}},~\quad\quad
\mathcal{R}_{12}=\sqrt{1-\beta_1^{(1)}}\sqrt{1-\beta_2^{(2)}},
\nonumber\\
\mathcal{R}_{20}&=&\sqrt{1-\beta_2^{(1)}}\sqrt{(1-\beta_1^{(2)})(1-\beta_2^{(2)})},~\quad\quad
\mathcal{R}_{22}=\sqrt{1-\beta_2^{(1)}}\sqrt{1-\beta_2^{(2)}}.
\end{eqnarray*}

\end{enumerate}

\section{Entanglement}

To quantify the  survival degree of entanglement $\mathcal{E}$
contained in the system, we use the negativity as a measure.  The
negativity  is defined as,
\begin{equation}
\mathcal{E}=max\left(0,\sum_i{\lambda_i}\right),
\end{equation}
where $\lambda_i,$ are the eigenvalues of the partial transpose of
$\rho^{T_a}_{ab}$\cite{Karpat}.

\begin{figure}
  \begin{center}
\includegraphics[width=19pc,height=19pc]{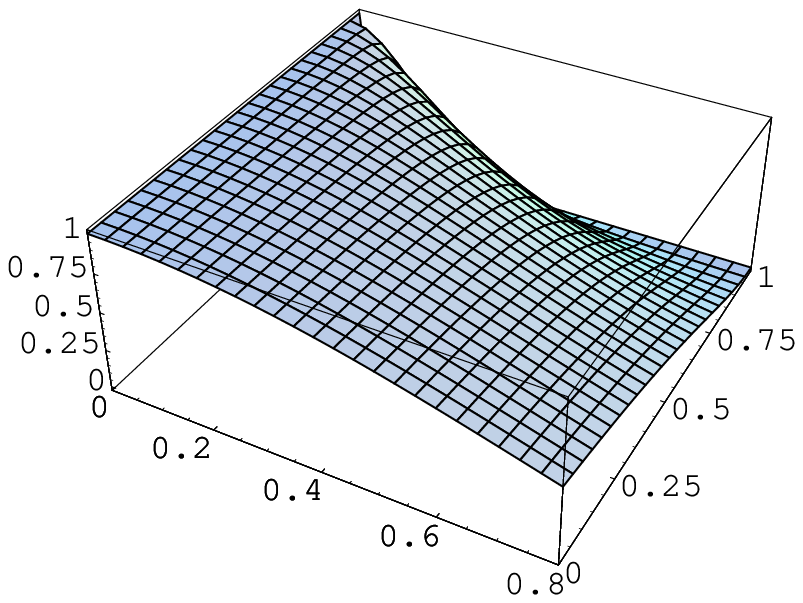}~~\quad
       \put(-160,35){\Large$ r$}
     \put(-240,110){\Large$\mathcal{E}$}
     \put(-20,70){\Large$\alpha_q$}
     \put(-60,210){$(a)$}~~\quad
      \includegraphics[width=19pc,height=19pc]{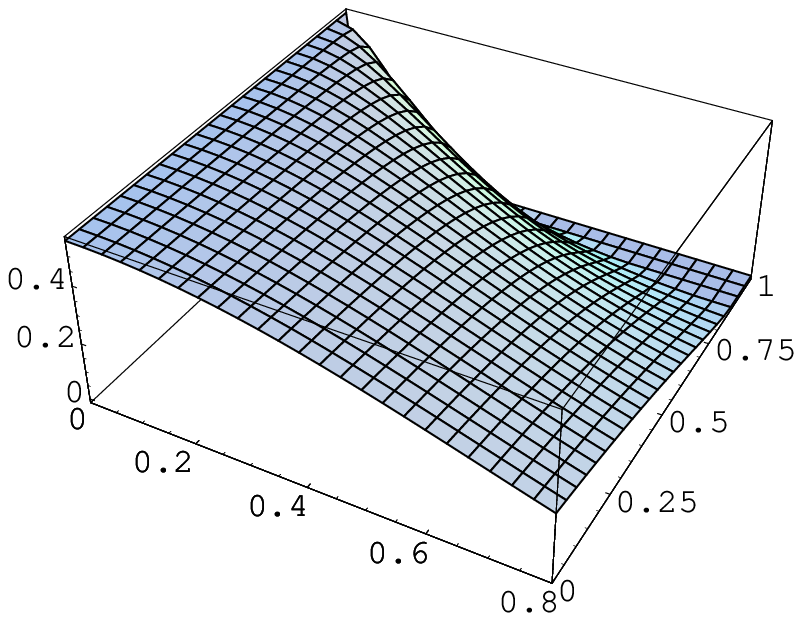}
  \put(-160,35){\Large$r$}
  \put(-235,100){\Large$\mathcal{E}$}
  \put(-15,70){\Large$\alpha_q$}
  \put(-60,210){$(b)$}
         \caption{The entanglement of the final state of the two-qubit system (8)
          where we set $\alpha_q^{(1)}=\alpha_q^{(2)}= \beta_q^{(1)}=\beta_q^{(2)}=\alpha_q$.
          (a) The initial state is prepared in a maximum entangled states,
          i.e., $(c_{11}=c_{22}=c_{33}=1$).
          (b) The initial state is prepared in a partial entangled state (Werner state, $c_{11}=c_{22}=c_{33}=0.7$),
           }
  \end{center}
\end{figure}

 Fig.(1a),  displays the  behavior  of entanglement of the final state (\ref{Final-qubit}), where
  it is assumed that, the partners share  initially a singlet state
i.e., $c_{11}=c_{22}=c_{33}=1$ and we  have set
$\alpha_q^{(1)}=\alpha_q^{(2)}=\beta_q^{(1)}=\beta_q^{(2)}=\alpha_q$.
 It is clear that, for  $r=0$ the degree of entanglement is maximum i.e.,
($\mathcal{E}=1)$. The general behavior shows that, the
entanglement decreases as the acceleration parameter, $r$
increases. On the other hand, as one increases the strengths of
the local operations, the entanglement increases. However, for
further values of $\alpha\in[0.75,1]$, the entanglement decreases.

The behavior of entanglement of a system  initially prepared in a
partially entangled state of Werner type is  displayed in
Fig.(1b). The general behavior is similar to that for the MES
(Fig.(1a)), but  with smaller  decay rate.  On the other hand, for
$\alpha\in[0.75,1]$, the entanglement re-appears again while, it
vanishes completely for the MES.

\begin{figure}[t!]
  \begin{center}
    \includegraphics[width=19pc,height=19pc]{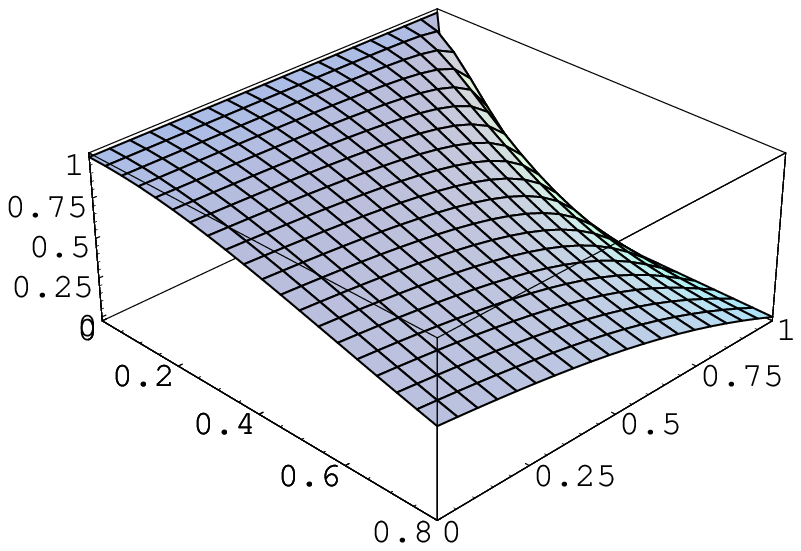}~~\quad
     \put(-180,35){\Large$ r$}
     \put(-240,130){\Large$\mathcal{E}$}
     \put(-25,35){\Large$\alpha_t$}
     \put(-60,210){$(a)$}~~\quad
    \includegraphics[width=19pc,height=19pc]{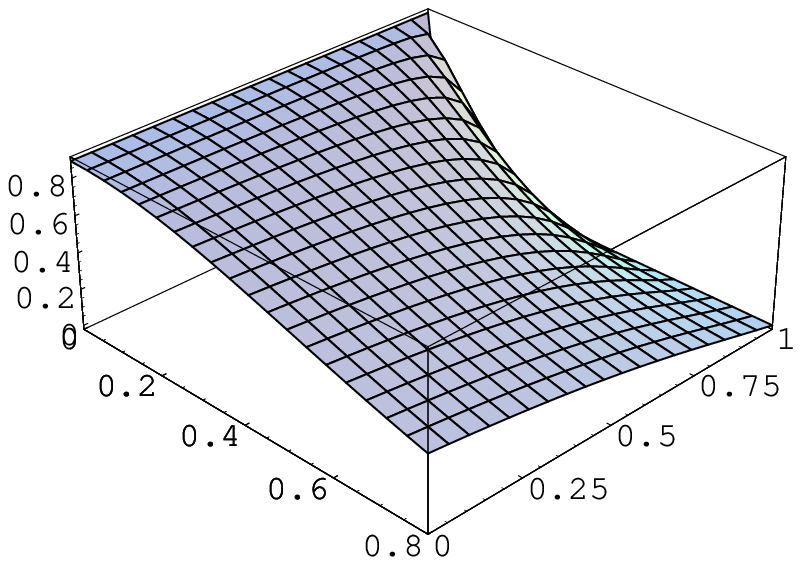}
 \put(-190,35){\Large$r$}
  \put(-235,130){\Large$\mathcal{E}$}
  \put(-30,40){\Large$\alpha_t$}
  \put(-60,210){$(b)$}
         \caption{The  of entanglement, of the two-qutrit system Eq.(11), where (a) maxim entangled state ($\gamma=1$)
         and (b) partial entangled state i.e., ($\gamma=0.5$), where set
         $\alpha_t^{(1)}=\alpha_t^{(2)}= \beta_t^{(1)}=\beta_t^{(2)}=\alpha_t$. }
  \end{center}
\end{figure}

The dynamics of entanglement under the effect of the local
weak-reverse measurement for a system is  initially prepared in a
two-qutrit system is described in Fig.(2).  The general behavior
is similar to that predicted in Figs.$(1)$, where the entanglement
decays as $r$ increases. However, the upper and lower bounds of
entanglement depend on initial state settings. It is clear that,
for the two-qutrit system the entanglement increases slowly
compared with that shown in Fig.(1).  Also, as displayed in
Fig.(2b), the increasing rate of entanglement for PES is larger
than that shown for MES.

From the previous figures, one  concludes that the weak and
reverse measurements can recover the loss of entanglement  for
small values of accelerations as the local operations' strengths
increase for different intervals. The length of these intervals
depends on the initial states settings: two-qubit, or two qutrits,
maximum / partial entangled states. The increasing rate of
entanglement is much larger for systems that are initially
prepared in partial entangled states. The partially entangled 2-
qubits state is more robust than the partially entangled
two-qutrit state.

\section{Local and Non-local Information}
  \subsection{Two-qubit system}
 In this subsection, we  quantify Alice's and Bob's information, which represents the
accelerated $(\mathcal{I}_a)$ and
non-accelerated,($\mathcal{I}_b$) information, respectively.
Moreover,  the amount of the non-local information between the
partners  defined by the coherent information
$(\mathcal{I}_{coh})$ will be quantified also. In an explicit
form, theses three types of information can be written as,

\begin{eqnarray}
\mathcal{I}_{a}&=&-\frac{\tilde\mathcal{B}_1+\tilde\mathcal{B}_3}{\mathcal{N}_{q}}\log\left(\frac{\tilde\mathcal{B}_1+\tilde\mathcal{B}_3}{\mathcal{N}_{q}}\right)-
\frac{\tilde\mathcal{B}_5+\tilde\mathcal{B}_7}{\mathcal{N}_{q}}\log\left(\frac{\tilde\mathcal{B}_5+\tilde\mathcal{B}_7}{\mathcal{N}_{q}}\right)
\nonumber\\
\mathcal{I}_{b}&=&-\frac{\tilde\mathcal{B}_1+\tilde\mathcal{B}_5}{\mathcal{N}_{q}}\log\left(\frac{\tilde\mathcal{B}_1+\tilde\mathcal{B}_5}{\mathcal{N}_q}\right)-
\frac{\tilde\mathcal{B}_3+\tilde\mathcal{B}_7}{\mathcal{N}_{q}}\log\left(\frac{\tilde\mathcal{B}_3+\tilde\mathcal{B}_7}{\mathcal{N}_{q}}\right)
\nonumber\\
I_{coh}&=&\sum_{i=1}^{4}\mu_i\log(\mu_i),
\end{eqnarray}
where
\begin{eqnarray*}
 \mu_{1,2}&=&\frac{1}{2\mathcal{N}_{q}}\Bigl\{(\tilde\mathcal{B}_1+\tilde\mathcal{B}_7)
\pm\sqrt{(\tilde\mathcal{B}_1-\tilde\mathcal{B}_7)^2+4\tilde\mathcal{B}_2\tilde\mathcal{B}_8}\Bigr\},
\nonumber\\
\mu_{3,4}&=&\frac{1}{2\mathcal{N}_{q}}\Bigl\{(\tilde\mathcal{B}_3+\tilde\mathcal{B}_5)
\pm\sqrt{(\tilde\mathcal{B}_3-\tilde\mathcal{B}_5)^2+4\tilde\mathcal{B}_4\tilde\mathcal{B}_6}\Bigr\},
\end{eqnarray*}

\begin{figure}
  \begin{center}
    \includegraphics[width=19pc,height=19pc]{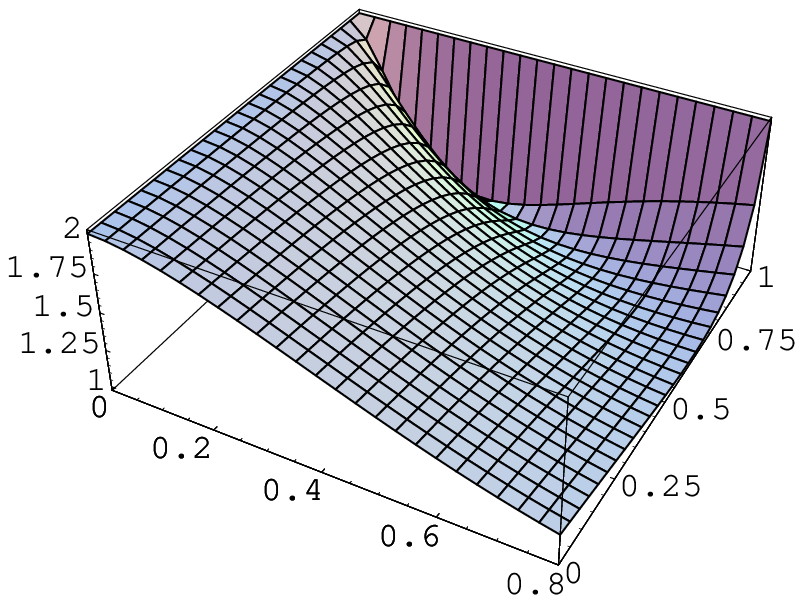}~~\quad
    \put(-160,30){\Large$r$}
     \put(-255,100){$\mathcal{I}_{coh}$}
     \put(-20,60){\Large$\alpha_q$}
      \put(-60,200){$(a)$}~\quad
    \includegraphics[width=19pc,height=19pc]{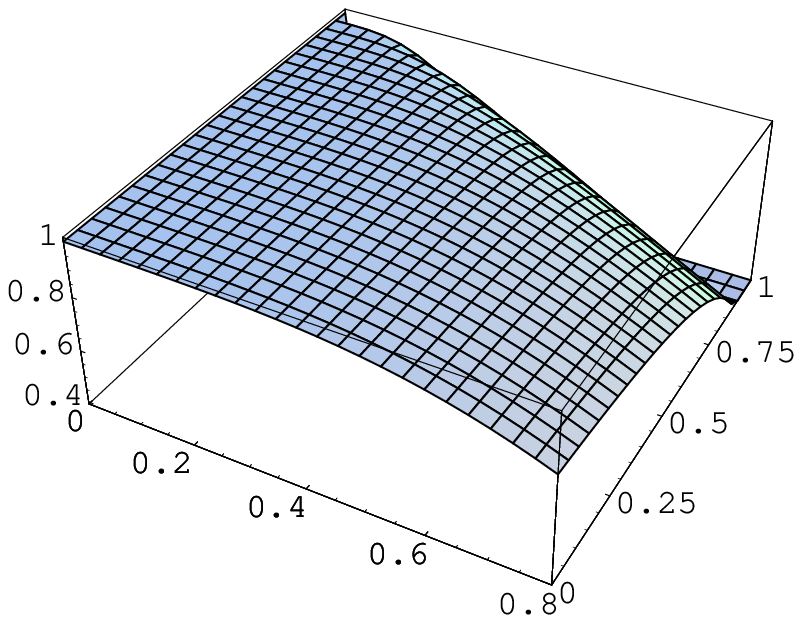}
 \put(-160,30){\Large$r$}
  \put(-235,97){$\mathcal{I}_a$}
  \put(-15,60){\Large$\alpha_q$}
   \put(-60,200){$(b)$}
         \caption{The local and non-local information for a system is initially
         prepared in a MES of two-qutrit system.
         (a) The coherent information between Alice and Bob, $\mathcal{I}_{coh}$ and (b) the accelerated local information
         $\mathcal{I}_a$. }
  \end{center}
\end{figure}

Fig.(3) displays the behavior of the coherent information and the
accelerated  local information, where we set
$\alpha_q^{(1)}=\alpha_q^{(2)}=
\beta_q^{(1)}=\beta_q^{(2)}=\alpha_q$. At zero acceleration, both
types of information are maximum. As Alice's particle is
accelerated, the coherent information decreases. However, for a
fixed value of $r$, the coherent information increases as the
strengths of the local operations increase. Meanwhile, the
accelerated local information increases at the expense of the
coherent information. For larger values of $\alpha_q$, the
accelerated information vanishes while the coherent information
increases. This coherent information, no longer represents quantum
information but it describes a classical information, because the
two particle are almost separable for larger values of $\alpha_q$
as displayed in Fig.(1).

 In Fig.(4), we scrutinize  the effect of the WRM on the
local and non local information, as well as on the coherent
information. The behavior of the accelerated information
$\mathcal{I}_a$ and the non-accelerated information
$\mathcal{I}_b$ are described in Fig.(4a), where it is assumed
that, the system is either initially prepared in a maximum
entangled state (MES) or in a  partial entangled state (PES). At
zero acceleration $(r=0)$ the information which is encoded in
Alice's and Bob's qubit is maximum i.e.
$\mathcal{I}_a=\mathcal{I}_b=1~ bit$. For small values of
$r\in[0,2.5]$, $I_a$ and $\mathcal{I}_b$ are remaining  maximum
$(1~bit)$. For larger values of $r$, the accelerated information
$\mathcal{I}_a$ decreases as the acceleration $r$ increases, while
Bob's information (non-accelerated information) is slightly
affected due to the local WRM. Moreover, the decay rate of the
accelerated information for a system   prepared initially  in a
partial entangled state is smaller than that displayed for systems
that  are initially prepared in MES. This shows that, the
possibility of protecting the accelerated local information
encoded in PES by the WRM is much better than protecting that
coded in MES.
\begin{figure}
  \begin{center}
    \includegraphics[width=20pc,height=15pc]{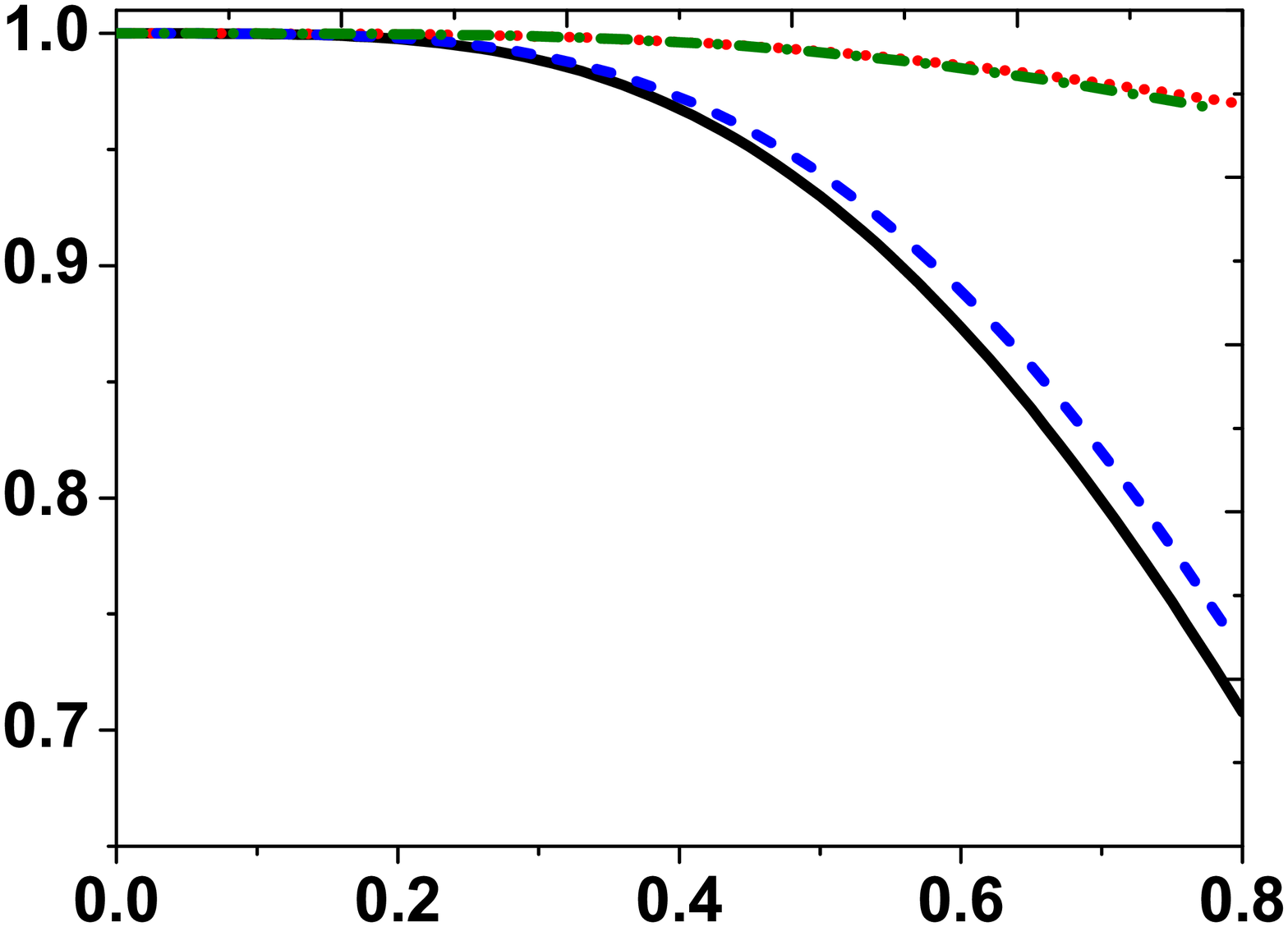}~\quad
        \put(-80,140){\Large$\mathcal{I}_{b}$}
  \put(-80,90){\Large$\mathcal{I}_{a}$}
   \put(-240,90){$\mathcal{I}_{Local}$}
  \put(-110,5){\Large$r$}
   \put(-60,150){$(a)$}
    \includegraphics[width=20pc,height=15pc]{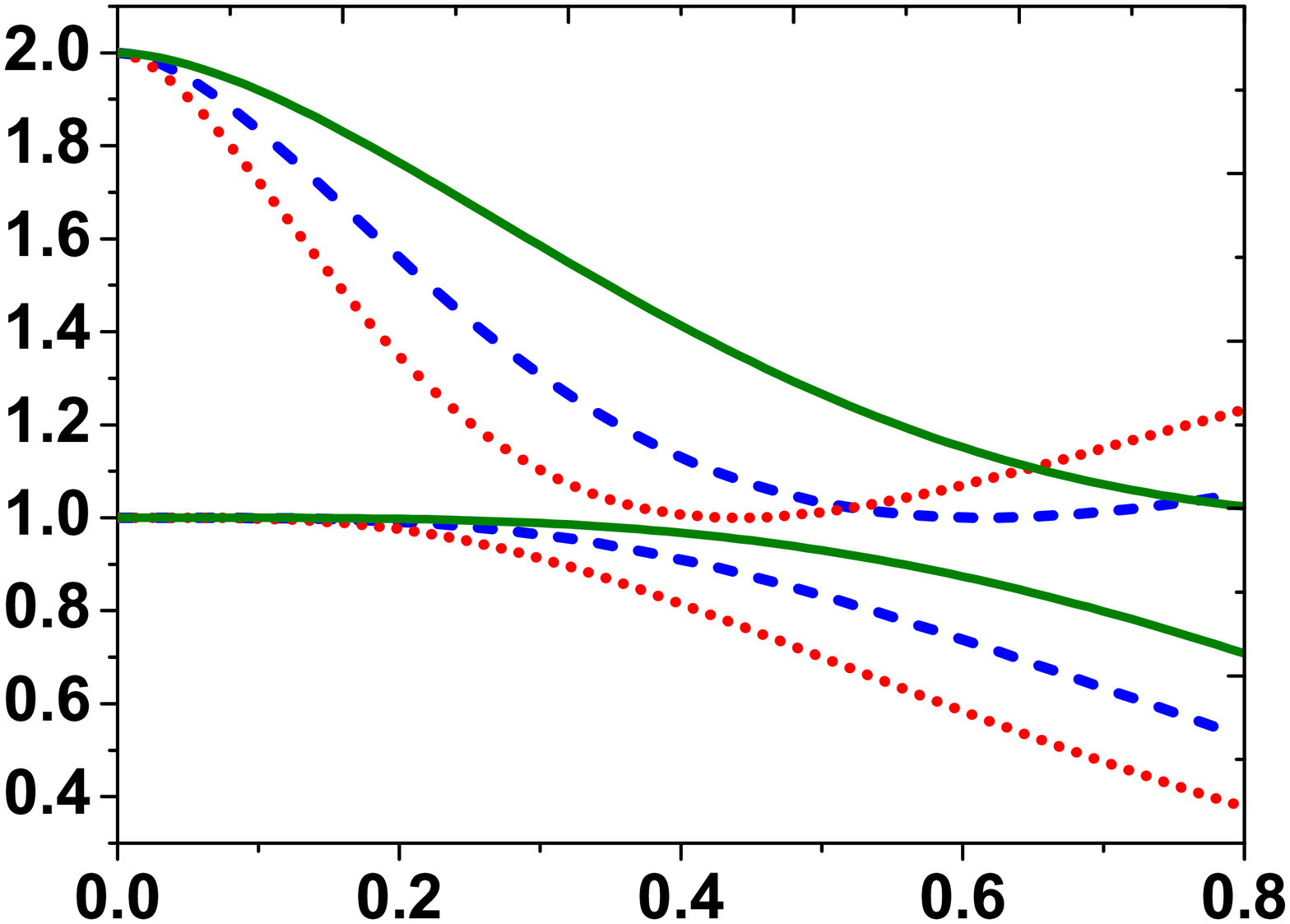}
  \put(-240,120){\Large$\mathcal{I}_{coh}$}
         \put(-240,60){\Large$\mathcal{I}_{a}$}
  \put(-110,5){\Large$r$}
   \put(-60,150){$(b)$}
         \caption{(a)Alice and Bob's information for a system is initially prepared in a
         MES or PES. The solid and dash curves represent Alice's
         for system is initially prepared in MES and PES,
         respectively, while the  dash-dot and dot curves represent
         Bob's local information for MES and PES respectively, where  we set
         $\alpha_q^{(1)}=\alpha_q^{(2)}= \beta_q^{(1)}=\beta_q^{(2)}=\alpha_q=0.5$
         (b) The local and non local information for different values of $\alpha_q=0.5,0.8,0.9$, for the solid,dash and dot curves
         respectively. }
  \end{center}
\end{figure}

The effect of the local WRM on the accelerated local information,
$\mathcal{I}_a$ and the coherent information $\mathcal{I}_{coh}$
is depicted in Fig.(4b), where it is assumed that the system  is
initially prepared in the MES. In these calculations, we consider
that the strengths of the local weak measurement and the reverse
measurements are equal $(\alpha_q=0.5)$. The behavior of
information shows that, both types of the information are maximum
at $r=0$. As $r$ increases, the accelerated local information
$\mathcal{I}_a$ has maximum values for $r\in[0,2.5]$, while the
coherent information decreases for any value of $r>0$. It is clear
that, as the strengths of the WRM increase, both types of
information decrease. However the decay rate of
$\mathcal{I}_{coh}$ is much larger than that for accelerated local
information, $\mathcal{I}_a$. For larger values of  $r$ and
$\alpha_q$, the coherent information increases at the expense of
the local accelerated  information.

From these figures, one  concludes  that the accelerated
information for small values of acceleration can be  protected by
using the weak-reverse measurement. The possibility of protecting
the accelerated information which is encoded in partial entangled
states is larger than that encoded in a maximum entangled states.
The local measurement has a very slight effect on the
non-accelerated local information. For larger values of
accelerations, the coherent information increases at the expense
of  the accelerated information

\subsection{Two-qutrit system}

Fig.(5) shows the behavior of the coherent information
$\mathcal{I}_{coh}$ and the accelerated information
$\mathcal{I}_a$ for a system  initially prepared in an entangled
maximum two-qutrit system. It is clear that, both types of
information decrease as the acceleration $r$ increases. The decay
rate is larger for the coherent information compared with that
depicted for the accelerated information. However, as the
strengths of the local operations increase  both types of
information are improved. For $\alpha_t\in[0.8,~1]$, the coherent
information increases at the expense of the  accelerated
information.
\begin{figure}
  \begin{center}
    \includegraphics[width=19pc,height=19pc]{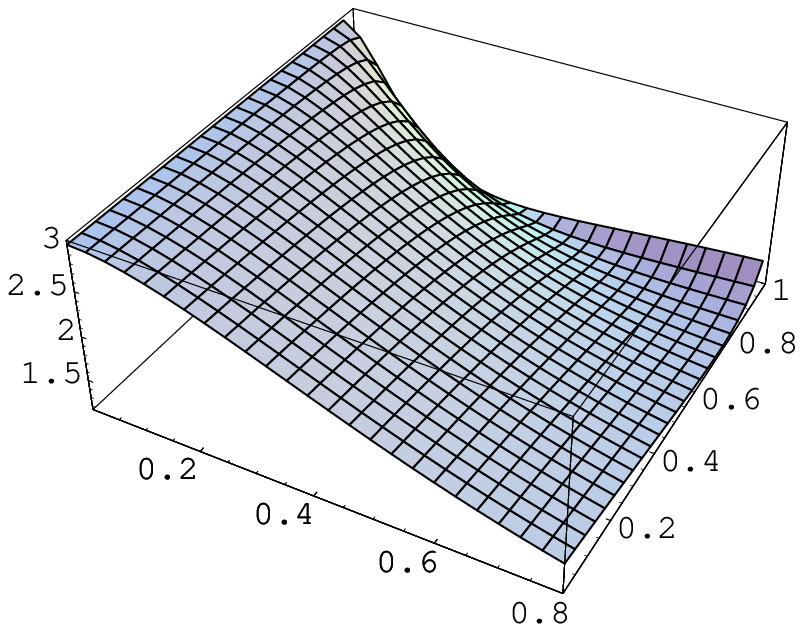}~~\quad
    \put(-160,28){\Large$r$}
     \put(-255,100){$\mathcal{I}_{coh}$}
     \put(-20,60){\Large$\alpha_t$}
      \put(-60,205){$(a)$}
    \includegraphics[width=19pc,height=19pc]{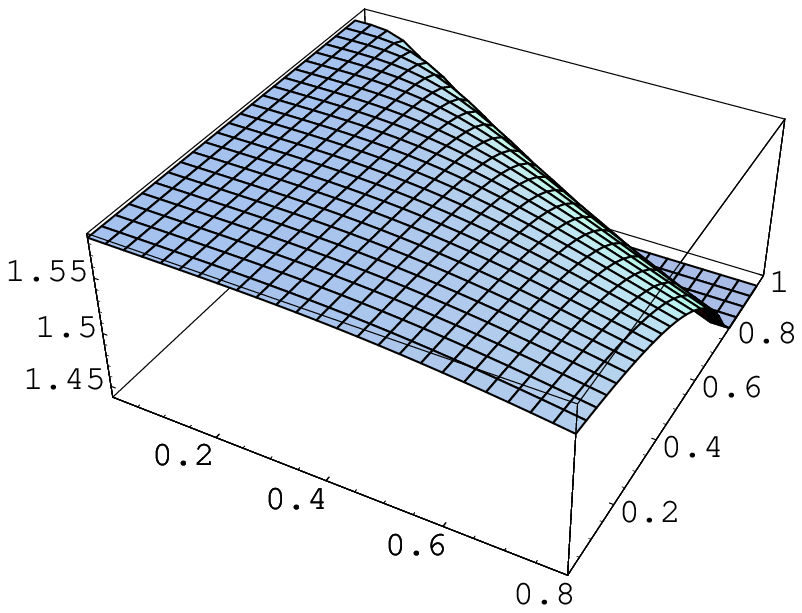}
 \put(-160,30){\Large$r$}
  \put(-230,98){$\mathcal{I}_a$}
  \put(-15,60){\Large$\alpha_t$}
   \put(-60,205){$(b)$}
         \caption{The same as Fig.(3), but the system is initially prepared in a maximum entangled 2-qutrits state. }
  \end{center}
\end{figure}

The dynamics of  different types of information coded in a system
 initially prepared in a maximum two- qutrit entangled state
is described in Fig.(6). The behavior of the accelerated local
information $\mathcal{I}_a$ and the non-accelerated local
information, $\mathcal{I}_b$ is displayed in Fig.(6a), where  it
is assumed that, the system  initially prepared in  MES or PES. It
is clear that, the accelerated information for a system is
initially prepared in a MES is  almost stable  and it is  a
maximum   for  any value of $r$. On the other hand, starting from
PES, the initial values of the accelerated information is smaller
than that depicted for MES. However,  the upper bounds of
$\mathcal{I}_a$ increase as the acceleration $r$ increases. These
upper bounds become larger than that depicted for MES for
$r\in[0,~0.8]$. For non-accelerated information the behavior is
completely different from those shown for the two-qubit system,
where it is always smaller than the accelerated information.
Moreover, it decreases as $r$ increases.

Fig(6b) shows the behavior of the coherent information and the
accelerated information for different values of the local
operation's strengths, with initial system  prepared in a MES of
the two-qutrit system.  It is clear that, $\mathcal{I}_{coh}$
decreases as $\alpha_t$ and $r$  increase. For smaller values of
$\alpha_t$, the coherent information decreases faster than that
displayed for larger values of $\alpha_t$.  On the other hand, for
larger values of $\alpha_t$ and $r$, the coherent information
increases to reach its upper bounds as $r\to \infty$. Also, the
general behavior of the  accelerate information  shows that,
$\mathcal{I}_a$ decreases as the strengths of the local operations
increases. However, the decay rate is smaller than the coherent
information. This explain the decay of the non-accelerated
information  displayed in Fig.(6a).

\begin{figure}[t!]
  \begin{center}
   \includegraphics[width=20pc,height=15pc]{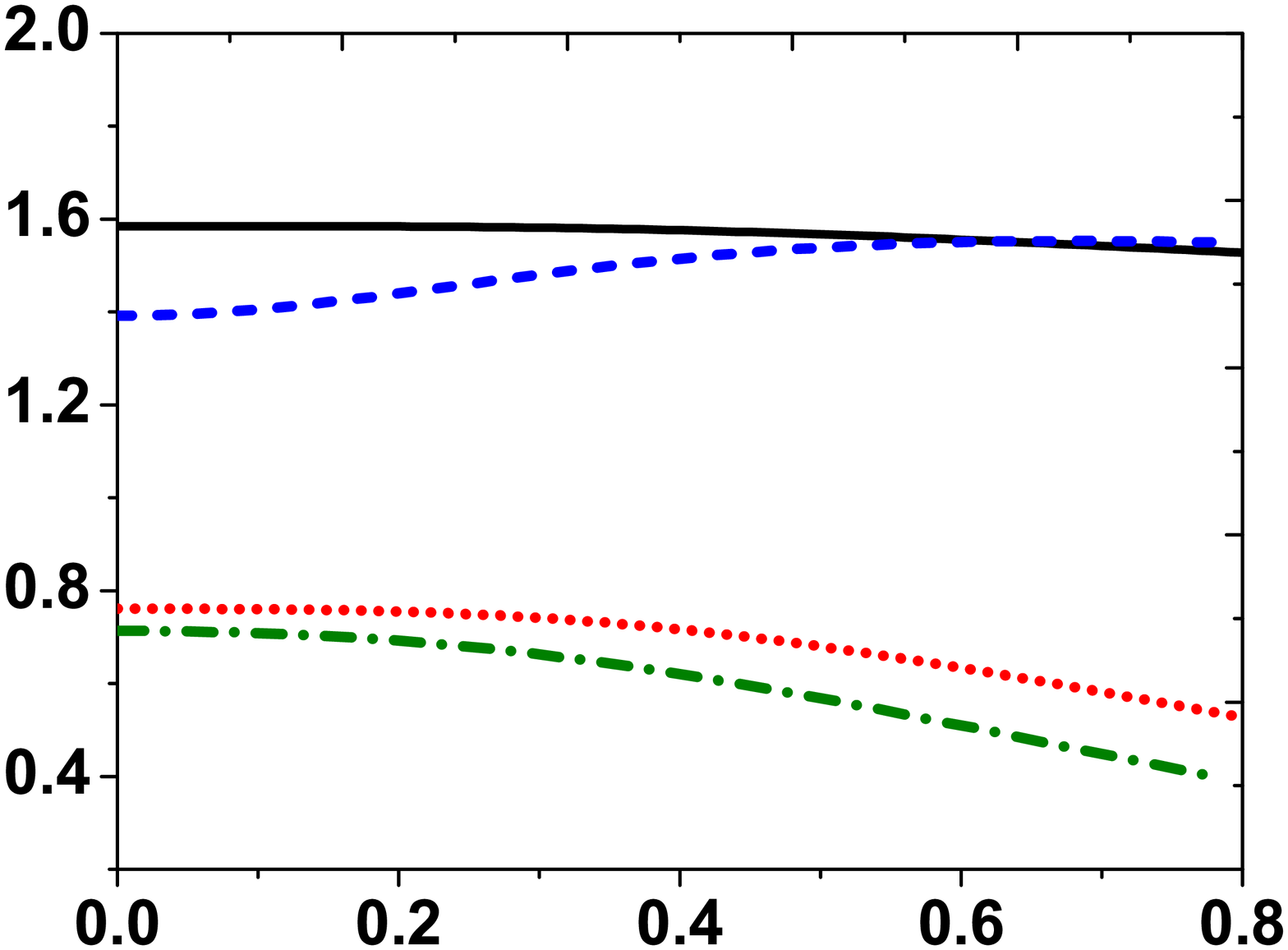}
 \put(-240,120){\Large$\mathcal{I}_{a}$}
  \put(-240,60){\Large$\mathcal{I}_{b}$}
  \put(-110,5){\Large$r$}
  \put(-60,150){$(a)$}
    \includegraphics[width=20pc,height=15pc]{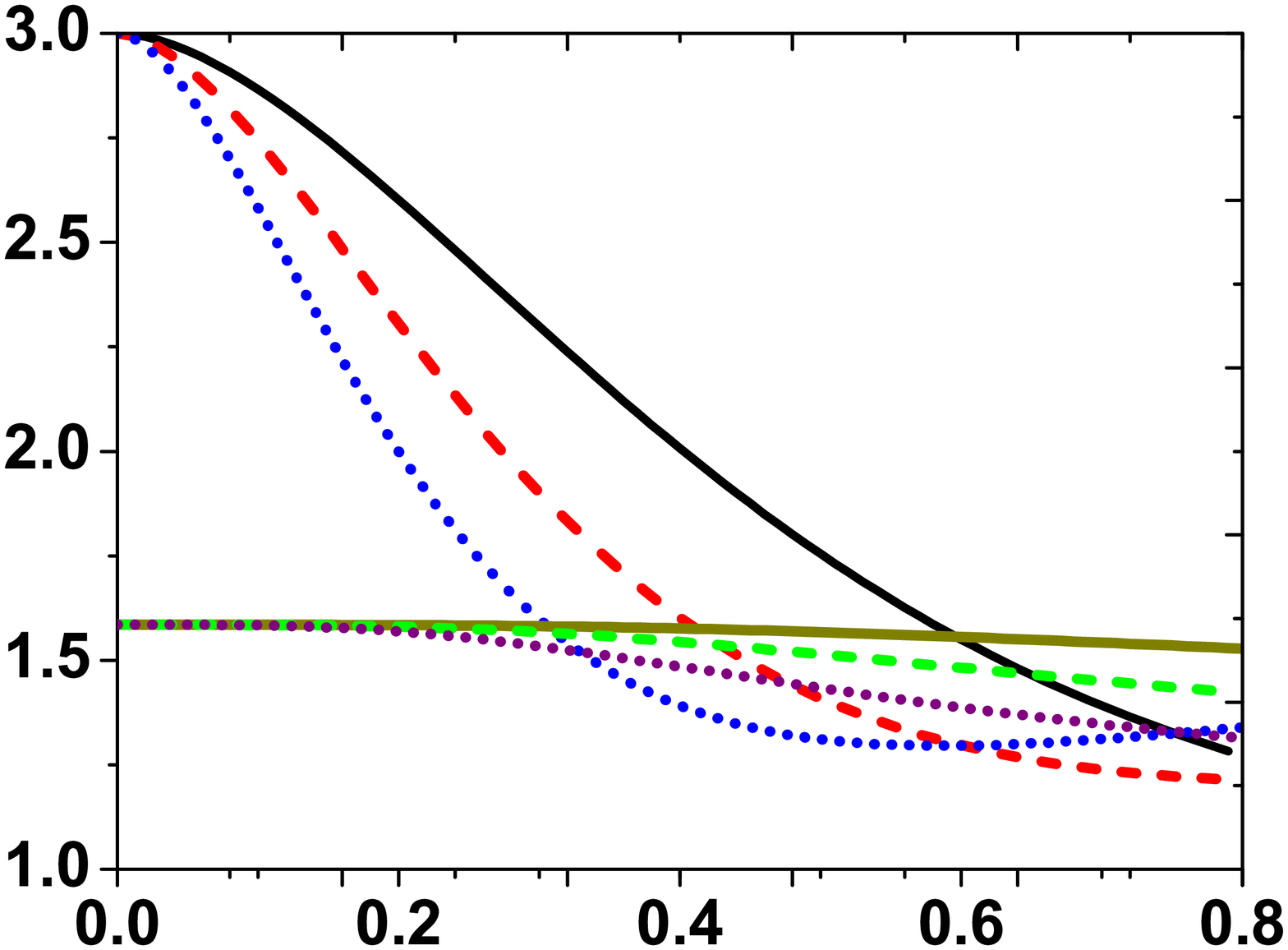}
        \put(-240,120){\Large$\mathcal{I}_{coh}$}
         \put(-240,60){\Large$\mathcal{I}_{a}$}
  \put(-110,5){\Large$r$}
   \put(-60,150){$(b)$}
            \caption{ Local and non-local information of a  MES  of two-qutrit system where $\alpha_t=0.5,0.8,0.9$. }
  \end{center}
\end{figure}

From these figures one can extract some important facts.  The
 local information which is  encoded in the accelerated
part of the composite system can be protect by using the WRM. The
local accelerated information which is encoded in a partial
entangled state can be improved even for larger accelerations. The
non-accelerated local information is very sensitive to the local
operations (WRM), compared with that shown for the two-qubit
system. The decay rate of accelerated information is much smaller
than that shown for the coherent information. For larger values of
accelerations, the coherent information increases at the expense
of the non-accelerated local information.

\section{Conclusion}

In this contribution, we assume that two users, Alice and Bob,
share 2-qubit or 2-qutrit states as a communication channels to
perform some quantum information tasks. Alice is   allowed to
accelerate her qubit/qutrit. Subsequently, the coherence of the
communication channel between the partners decreases. Thus, we try
to investigate the possibility of utilizing the weak-reverse
technique to improve the quantum correlations between the users.
It is shown that, the initial state settings, the values of the
acceleration and the strengths of the weak-reverse operations
represent  control keys to the behavior of entanglement, local and
non-local information.

It is demonstrated that, for small values of acceleration the
improving rate of the quantum correlation of the communication
channels increases as the strengths of the weak-reverse operation
increase. However, as one increases the local operations'
strengths further the correlation decreases. Our results show
that, starting from less entangled state the improving rate of
correlations is better than that depicted for maximum entangled
states. On the other hand, the increasing rate of the correlation
for the two-qutrit system is larger for large values of the
acceleration comparing with that displayed for the
two-qubit-system.

Additionally,  the behavior of the local information which
includes the  accelerated and non-accelerated information is
assayed. It is shown that, for  the two-qubit system, the
non-accelerated information is almost stable and slightly
decreases for systems  initially prepared in maximum entangled
state. But the accelerated information decreases as the
acceleration increases. Moreover,  the encoded information in a
system  initially prepared in partial entangled state is more
robust than that encoded in a system  initially prepared in
maximum entangled states. Theses results are dramatically changed
for the two-qutrit systems, where the non- accelerated information
is more fragile than the accelerated information. Also, the
encoded information in the partial entangled state of the
two-qutrit system is improved as the acceleration increases under
weak-reverse measurements.

Finally, the general  behavior of the coherent information
displays that. it decreases as the acceleration increases. For  a
fixed value of acceleration, the coherent information increases as
the strength of the weak-reverse measurement increases. However,
for the two-qubit system, larger values of these strengths cause a
sudden decay of information for small values of the accelerations
and gradually decay for larger values of these accelerations.
Meanwhile, for the two-qutrit system, the coherent information
slightly increases  at larger values of the weak-reverse
measurements' strengths. The changes  happened at the expense of
the local information.

{\it In conclusion}, it is possible to improve and protect the
accelerated communication channel by using weak-reverse
measurements. The possibility of protecting the local encoded
information on an accelerated 2-qutrit system is larger than that
displayed for 2-qubit system. The restrained decay rate  of the
coded information on accelerated partial entangled state is better
than that displayed for maximum entangled state. The encoded
information in the accelerated part of the two-qutrit system can
be improved by using the local operations, while it decays for the
two-qubit system. Therefore, to perform quantum key distribution
or quantum coding protocols  using accelerated system, it is
better to use partial entangled 2-qutrit systems with larger
accelerations.


\begin{thebibliography}{99}

\bibitem{Bennt}C.H. Bennett, G. Brassard, S. Popescu, B. Schumacher, J.A. Smolin, W.K.
Wootters,"Purification of Noisy Entanglement and Faithful
Teleportation via Noisy Channels", Phys. Rev. Lett. {\bf 76} 722
(1996).
\bibitem{Deutsch} D. Deutsch, A. Ekert, R. Jozsa, C.
Macchiavello, S. Popescu, A. Sanpera,"Quantum Privacy
Amplification and the Security of Quantum Cryptography over Noisy
Channels", Phys. Rev. Lett. {\bf 77}
 2818 (1996).

 \bibitem{Metwally2002}N. Metwally,"More efficient entanglement purification",
  Phys. Rev. A {\bf 66}  054302 (2002).

  \bibitem{Metwally2006} N. Metwally, A.-S. Obada, "More efficient purifying scheme via controlled–controlled NOT
  gate", Physics Letters A {\bf 352}  45 (2006).

\bibitem{Moreno} H.. J. Moreno, T. Gorin and T. H. Seligman,"
Improving coherence with nested environments", Phys. Rev. A. {\bf
92} 030104(R) (2015).
\bibitem{Xiao2016} X. Xiao, Y. Yao, Wo-J. Zhong, Y.-L. Li and Y.M.
Xie," Enhancing teleportation of quantum Fisher information by
partial measurements" , Phys. Rev. A {\bf 93} 012307 (2016).

\bibitem{Yas2014} K. O. Yashodamma, P. J. Geetha and Sudha,"
Purification an redistribution of entanglement via single local
filtering", Int. J. Quantum Innfor. {\bf 12} 1450004 (2014).

\bibitem{Metwally2016-0} N. Metwally"Immunity, Improving and Retrieving the lost entanglement of accelerated qubit-qutrit
 system via local Filtering", h > arXiv:1603.01429 (2016).


\bibitem{Liang} J.-L. Guo, J.-L.Wi and W. Qin," Enhanemencent of
quantum correlations in qubit-qutrit system under decoherence of
finite temperature", Quantum In Process {\bf 14} 1399-1410 (2015).

\bibitem{Xiao} X.Xiao," Protecting qubit-qutrit entanglement from
amplitude damping decoherence from amplitude damping decoherence
via weak measurement and reversel", Phy. Scr. {\bf 89} 065102
(7pp) (2014).




\bibitem{Metwally2013} N. Metwally" Telelportation of
accelerated Information" J. Opt. Soci. Am B {\bf 30} 233 (2013).

\bibitem{Downes} T. G. Downes, T. C. Ralph and N. Walk," Quantum
communication with an accelerated partner", Phys. Rev. A {\bf 87}
012327 (2013).
\bibitem{Sagheer} N. Metwally and A. Sagheer" Quantum coding in
non-inertial frames", Quantum Information Processing {\bf 13} 771
(2014).

\bibitem{Metwally2016} N. Metwally," Entanglement of simultaneous and non-simultaneous
accelerated qubit-qutrit systems", Quantum Inf. and Comput (QIC),
{\bf 16} 0530-0542 (2016).

\bibitem{kim} Y. S. Kim, J. C. Lee, O. Kwon, Y. H. Kim,"Protecting
entanglement from decoherence using weak measurement and quantum
measurement reversal", Nat Phys. {\bf 8} 117 (2012).
\bibitem{Xiao2013} X. Xiao, T, L. Li "Protecting qutrit-qutrit
entanglement by weak measurement and reversal, Eur. Phys. J. D
{\bf 67} 204 (2013).

\bibitem{Jason2013}E. M.-Martinez, I. Fuentes,"Redistribution of particle and antiparticle entanglement in noninertial frames",
 Phys. Rev. A {\bf 83}, 052306
(2011); J.  Doukas, E. G. Brown, A. J. Dragan, and R. B. Mann,
"Entanglement and discord: Accelerated observations of local and
global modes", Phys. Rev. A {\bf 87}, 012306 (2013).

\bibitem{MetwallyJMPB} N. Metwally, J. Mod. Phys. B {\bf 27} 1350155
(18pages) (2013).
\bibitem{Parisio} E. A. Fonseca and F. Parisio," Maesure of nonlocality which is maximal for maximally
entangled qutrits", Phys. Rev. A {\bf 92} 030101 (R) (2015).

\bibitem{Karpat} G. Karpat and Z. Gedik, Phys. Lett. A {\bf 375}
4166-4171 (2011).


\end{thebibliography}
\end{document}